\documentclass[sigplan,10pt]{acmart}
\AtBeginDocument{%
  \providecommand\BibTeX{{%
    \normalfont B\kern-0.5em{\scshape i\kern-0.25em b}\kern-0.8em\TeX}}}

\setcopyright{acmcopyright}
\copyrightyear{2018}
\acmYear{2018}
\acmDOI{XXXXXXX.XXXXXXX}

\acmConference[Conference acronym 'XX]{Make sure to enter the correct
  conference title from your rights confirmation emai}{June 03--05,
  2018}{Woodstock, NY}
\usepackage{listings}
\usepackage{multirow}
\usepackage{pdfpages}
\usepackage{graphicx}
\usepackage{caption}
\usepackage{pgf-pie}
\usepackage{pgfplots}
\usepackage{algorithmic}
\usepackage{algorithm}
\pgfplotsset{compat=1.17}
\usepackage{subcaption}
\usepackage{float}
\usepackage{adjustbox}
\usepackage{subcaption}
\usepackage{cleveref}

\usepackage[textsize=scriptsize,textwidth=1.2cm]{todonotes}

\begin{document}

\title{Can ChatGPT Pass An Introductory Level Functional Language Programming Course?}

\author{Chuqin Geng}
\email{chuqin.geng@mail.mcgill.ca}
\affiliation{%
  \institution{McGill University}
  \city{Montreal}
  \state{Quebec}
  \country{Canada}
}

\author{Yihan Zhang}
\email{yihan.zhang2@mail.mcgill.ca}
\affiliation{%
  \institution{McGill University}
  \city{Montreal}
  \state{Quebec}
  \country{Canada}
}

\author{Brigitte Pientka}
\email{bpientka@cs.mcgill.ca}
\affiliation{%
  \institution{McGill University}
  \city{Montreal}
  \state{Quebec}
  \country{Canada}
}

\author{Xujie Si}
\email{six@cs.toronto.edu}
\affiliation{%
 \institution{University of Toronto}
 \city{Toronto}
 \state{Ontario}
 \country{Canada}}




\renewcommand{\shortauthors}{Trovato and Tobin, et al.}

\begin{abstract}

The recent introduction of ChatGPT has drawn significant attention
from both industry and academia due to its impressive capabilities in
solving a diverse range of tasks, including language translation, text
summarization, and computer programming.
Its capability for writing, modifying, and even
correcting code together with its ease of use and access is already 
dramatically impacting computer science education. 
This paper aims to explore how well ChatGPT can perform in an
introductory-level functional language programming course. In our
systematic evaluation, we treated ChatGPT as one of our students and
demonstrated that it can achieve a grade B- and its rank in the class is 155 out of
314 students overall. Our comprehensive evaluation provides valuable
insights into ChatGPT's impact from both student and instructor
perspectives. Additionally, we identify several potential benefits
that ChatGPT can offer to both groups. Overall, we believe that this
study significantly clarifies and advances our understanding of
ChatGPT's capabilities and potential impact on computer science education.


\end{abstract}






\maketitle

\section{Introduction}
Software systems based on artificial intelligence (AI) have shown great potential in computer science education. They range from intelligent tutoring systems \cite{crow2018intelligent,abu2008developing,nwana1990intelligent} to automated grading systems \cite{aldriye2019automated, lee2021effectiveness,edwards2008web} and plagiarism detection tools \cite{devore2020mossad,al2010plagdetect,hage2010comparison,bowyer1999experience}. Most recently, the easy access to and the power of large language models is transforming education.  Models such as GPT-3 \cite{GPT-3} and its variants seemingly understand and respond to user input in a conversational manner based on the given prompt. Users can modify and fine-tune the prompt through a process called prompt engineering, which better instructs the model to complete different tasks such as answering basic questions and providing more complex explanations and solutions 
\cite{zhou2022large, strobelt2022interactive, white2023prompt}. 
%
%
%
%
%
%
One of the most notable large language models is ChatGPT\footnote{https://openai.com/blog/chatgpt/ (accessed March 11, 2023).}, which has garnered significant attention due to its ability to engage in conversational interactions with users and incorporate previous chat history into its responses \cite{liu2023summary}. 
While initial experiments with using ChatGPT to write source code and provide suggestions for improving code quality, have been promising \cite{tian2023chatgpt}, it remains unclear how well ChatGPT performs on actual problem sets and exams in a computer science classroom. There are many questions that remain open: Can ChatGPT accurately understand and interpret the task at hand?  Can it deal with complex coding assignments and use specific and advanced programming concepts? Is the generated solution sensible, correct, and efficient? 

In this study, we evaluate ChatGPT's performance in an introductory functional language programming course that focuses on OCaml based on all homework assignments and exams from the 2022 fall semester.
Our goal is  to assess the grade that ChatGPT would receive on each assessment as well as overall in the course. To compare the model's performance to actual students enrolled in the course, we also rank its performance with respect to actual grade distributions in the course. 

\begin{figure}
    \centering    \includegraphics[width=.8\linewidth,trim={5cm 2cm 7cm 5cm}, clip]{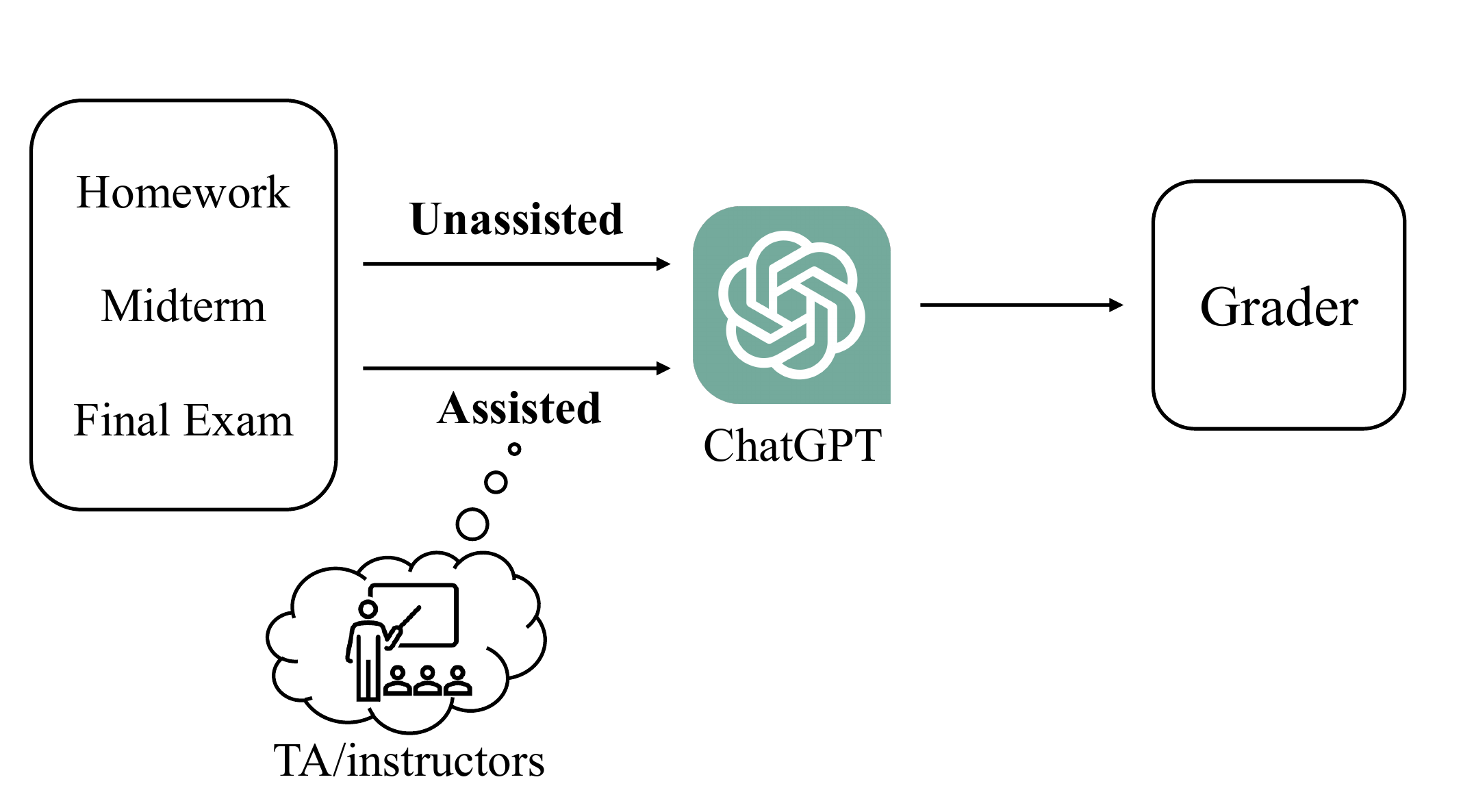}
    \caption{A flowchart illustrating our study design. In this study, we design two types of experiment which are \emph{Unassisted} and \emph{Assisted}.}
    \label{fig:flowchart}
\end{figure}

To judge ChatGPT's abilities with solving homework problems and exam questions, we use a two-pronged approach: we use ChatGPT \emph{unassisted} and \emph{assisted} (see Figure \ref{fig:flowchart}). In the unassisted mode, we use ChatGPT without prior training or knowledge engineering.
In the assisted mode, we facilitate ChatGPT's ability to complete specific tasks by providing it with necessary knowledge through prompt engineering. In particular, we designed and used four prompt engineering techniques: 1) paraphrasing the problem, 2) natural language hints, 3) teaching by example, and 4) providing test cases. 
These prompts that we give are similar to the assistance that a student might receive during office hours.
Our findings indicate that these techniques can significantly improve the quality of ChatGPT's response.
By comparing the results between the unassisted and assisted setting, we can determine the extent to which the quality of ChatGPT's answers can be improved through guided learning.

Our results show that ChatGPT achieves overall grade of 67\% (B-) in the course, which is slightly lower than the actual course average in fall 2022. ChatGPT obtains 100\% on 16 out of 31 assignment problems without any additional help. This is impressive as some of the questions require a lot of background knowledge. What is more, ChatGPT is particularly good at applying higher-order functions in OCaml, which is considered one of the most challenging parts for students, as it requires students to think abstractly over common functionality. On the other hand, ChatGPT presently still struggles with understanding type specifications and inferring the type of an expression. It also struggles with larger programming tasks such as implementing an interpreter. 
Furthermore, we validate the effectiveness of the prompting engineering approach by observing an increase in ChatGPT's ranking from 220th to 155th out of a total of 314 students. Overall, our results show the power and potential of solving programming tasks using ChatGPT. It also provides insights into how ChatGPT or similar tools can become a valuable tool for students and instructors. 

\section{Background}


In recent years, the development of large language models has accelerated with the advances in computing hardware such as GPU and TPU and the availability of vast amounts of training data. The success of these large language models is largely due to the transition in deep learning architecture from RNN (recurrent neural network) \cite{RNN} to Transformer models \cite{Transformer}. In fact, most successful large language models today such as BERT (Bidirectional Encoder Representations from Transformers) \cite{BERT} and the T5 (Text-to-Text Transfer Transformer) \cite{T5} are based on the Transformer. 


One of the most notable examples of this trend is the GPT (Generative Pre-trained Transformer) family of models developed by OpenAI. Early language models like GPT-1 \cite{GPT} had a relatively small number of parameters, but with the introduction of GPT-2 \cite{GPT-2} and GPT-3 \cite{GPT-3}, the size of language models has grown substantially. GPT-3, for instance, has 175 billion parameters, making it one of the largest language models to date. On the other hand, ChatGPT, an enhanced version of OpenAI's GPT-3, also referred to as GPT-3.5, has proven to outperform GPT-3 in almost every benchmark. With these large language models, we have seen significant improvements in natural language processing (NLP) tasks such as text generation, summarization, and translation. Additionally, the growing capabilities of large language models have enabled them to be used for various applications, including chatbots, virtual assistants, and even code generation. As the capabilities of these models continue to expand, we can expect to see further advancements in NLP and other areas of AI. 

The increasing popularity of large language models such as ChatGPT is due to their ability to enable users to  interact with them via only natural language texts. Such ability is deeply linked to their capability of predicting the next word given prior words in a sentence. Since most natural language-related tasks can be framed as the next-word prediction, ChatGPT as well as other GPT models are expected to perform various tasks if they are well-engineered as prompts/instructions. Thus, it is well-expected that ChatGPT could perform differently when different prompt engineering techniques are used to guide its responses. With the right prompt engineering, ChatGPT can be used to perform a wide range of tasks, such as language translation, summarization, and even code generation. As a result, ChatGPT has the potential to become a valuable tool in various domains, from education to healthcare, as it can help users complete tasks and solve problems using only natural language inputs.

When it comes to the impact of ChatGPT on computer science education, it is worth noting that the model's ability to understand and generate code represents a significant improvement over previous models. This breakthrough makes ChatGPT a tool that will transform computer science education.
As we have seen from previous work \cite{chen2021codex}, ChatGPT's predecessors
are capable of understanding and generating code in commonly-used programming languages such as Python, Java, and C. Thus, in this study, we focus on evaluating and analyzing ChatGPT's performance in understanding concepts and generating code in OCaml - a functional programming language, in a real university course teaching environment. Subsequently, based on our empirical results, we aim to investigate how both students and instructors/TAs can benefit from ChatGPT. While there are potential merits to using ChatGPT in computer science education, we also discuss its limitations and potential risks.

\section{Related Work}


Since large language models have shown great potential in code generation and repair tasks, there are many studies on the evaluation of large language models' capabilities on  a vast variety of programming tasks. For instance, AlphaCode \cite{alphacode}, proposed by DeepMind, is one of the first language model-based systems that demonstrated impressive performance in coding contests. In their work, the authors used a GPT-3 model to generate solutions to programming problems in Python. AlphaCode achieved a higher score than 45.7\% of human participants in a competition involving more than 5,000 participants. Similarly, GitHub has launched Copilot \cite{chen2021codex}, a GPT-3 based model that can assist programmers in completing code or even generating programs. Copilot uses a combination of supervised learning and unsupervised learning to generate code snippets based on natural language prompts, such as docstrings and function names or signatures. Several studies \cite{nguyen2022empirical,barke2023grounded,asare2022github,al2022readable} have evaluated the performance of Copilot on various programming tasks, and the results indicate that it can generate high-quality code with a low error rate. 

Another notable work in this field is CodeBERT \cite{Codebert_2020}, a bimodal pre-trained model for programming language and natural language that aims to improve downstream applications such as natural language code search and code documentation generation. The model is developed using an encoder of Transformer and is trained with a hybrid objective function that includes the pre-training task of replaced token detection. This approach allows CodeBERT to utilize both bimodal and unimodal data, where the former provides input tokens for training and the latter helps to learn better generators. Results from fine-tuning the model parameters show that CodeBERT achieves state-of-the-art performance on both natural language code search and code documentation generation. 

Unlike the previous models that leverage either the decoder or encoder of the Transformer model, CodeTrans \cite{CodeTrans} is an encoder-decoder transformer model designed for tasks in software engineering. Empirical studies show the effectiveness of this model in 6 different software engineering tasks, encompassing 13 sub-tasks. The paper also explores the effect of various training strategies, such as single-task learning, transfer learning, multi-task learning, and multi-task learning with fine-tuning. CodeTrans outperforms existing state-of-the-art models in all proposed software engineering tasks.

In addition to generating code, large language models have also been evaluated for code repair tasks. \citet{sobania2023analysis} evaluate the performance of ChatGPT on automated program repair, specifically on the QuixBugs benchmark set. The authors compare ChatGPT's bug-fixing performance with other standard program repair approaches and deep learning methods, such as CoCoNut and Codex. The results show that ChatGPT's performance is competitive with these approaches and significantly better than standard program repair approaches. Additionally, the convenience of the dialogue system of ChatGPT allows users to provide further information, resulting in a higher success rate of 31 out of 40 bugs fixed. The authors suggest that ChatGPT has the potential for automated program repair (APR) and can be further improved with additional user input. 

\citet{jiang2023impact} 
assess the effectiveness of large language models in program repair and investigate the potential of fine-tuning these models for the automated program repair (APR) task. The authors evaluate ten language models using four APR benchmarks, revealing that the best language models can fix 72\% more bugs than state-of-the-art deep-learning-based APR techniques without any modifications. To avoid data leaking, they also create a new benchmark and demonstrate that fine-tuning language models with APR training data lead to significant improvements in bug-fixing capabilities. The study also explores the impact of buggy lines and assesses the size, time, and memory efficiency of different language models. The paper also highlights potential areas for future research in the APR domain, such as fine-tuning language models with APR-specific designs, and emphasizes the importance of fair and comprehensive evaluations of language models.

Although these studies have conducted thorough evaluations and demonstrated the effectiveness of large language models in tasks such as code generation and repair, they have not been evaluated on coding tasks of a real programming course to better understand the impact on computer science education. 




\section{Study Design}


\subsection{Overview of The Course}
Our study concerns students in a second-year undergraduate computer science course at X\footnote{The name is intentionally anonymized for the double-blind review.} university. The course introduces concepts about functional programming and programming paradigms. It is offered every semester with more than 300 registered undergraduate students. In this study, we use assignments and exam questions from Fall 2022, since ChatGPT is trained on data before 2021, so it is less likely that the model has seen those questions before.

The course has ten weekly programming assignments each worth 3\%, a midterm exam worth 20\%, and a final exam worth 50\%   of the total grade. Table \ref{overview} presents a brief overview of the course content and related homework assignments. Each homework consists of several programming tasks to implement functions and test cases. All homework assignments were hosted on Learn-OCaml, an online programming platform for OCaml that allows students to edit, compile, test, and debug code all in one place. We used a modified version of Learn-OCaml \cite{10.1145/3341719} by Hameer and Pientka with additional features such as style checking and evaluation of test cases written by students.



\begin{table}[h!]
 \begin{adjustbox}{scale=0.8}
\begin{tabular}{|l | l | c | l |} 

 \hline
 Week & Topics & HW\# \  \\ [0.3ex] 
 \hline
 1-2 & Basics, recursion and pattern-matching & 1\\ 
 \hline
 3-4 & Higher-order function, exceptions and modules & 2\\ 
 \hline
 5 & Continuations & 3\\ 
 \hline
  6 & Reasoning about recursive algorithms & 4\\ 
 \hline
 7 & Mutable states and OOP  & 5\\ 
 \hline
 8 & Lazy programming & 6\\ 
 \hline
 9 & Fundamentals of PL theory & 7\\ 
  \hline
  10 & Substitution, evaluation and closures & 8\\ 
  \hline
 11-12 & Implementing a type system& 9\\ 
  \hline
13 & Defunctionalized continuations and stack machines & 10\\ 
  \hline
\end{tabular}
\end{adjustbox}
\caption{\label{overview}Overview of course schedule.}
\end{table}

\subsection{Design of Experiments}

Our dataset consists of the 2022 fall homework questions and exams, since ChatGPT is trained on data before 2021, so it is less likely that the model has seen those questions before. At the initial step of our experiment we focus on evaluation of solutions automatically produced by using ChatGPT, which we refer as \emph{unassisted experiments}. We view ChatGPT as a student in this course and ask it to complete all the ten assignments (total of 31 questions) from 2022 fall semester, covering a variety of topics including pattern-matching, recursion, backtracking, continuation and higher-order-functions. And the last two homework involves implementing a programming language called MiniOCaml in OCaml. The response code generated by ChatGPT is copy-pasted in the LearnOCaml platform to grade. We repeat the query five times and compute the average score to account for the heuristic nature of ChatGPT. The possible errors are classified into 3 categories: 1) \emph{OCaml syntax Errors}, which appear to be the most frequent mistakes for novice programmer in OCaml; 2) \emph{Compilation Errors}, which mainly refers to type mismatch errors; 3) \emph{Logical Errors}, which occurs when the solution fails to pass one or more test cases by the autograder even if the program compiles successfully. Based on our classification, each solution provided by ChatGPT can only be correct or has one of the error types above.

We also employ the same strategy for final exam and midterm in 2022 fall term. The exams are usually a mix of programming tasks and conceptual problems, which can be broken into 3 parts: Theory, Code Completion and Induction Proof. For each part, Theory contains conceptual questions related to type inference and OCaml fundamentals while the induction proof requires using mathematical induction to justify the correctness of programs. They are often multiple-choice questions. Code completion tasks asks students to write codes on paper or fill in blanks for an OCaml program. For fair comparison, all the exam answers returned by ChatGPT are manually graded by TA's in this course with the same rubrics as in 2022 Fall term.

We structure our procedure for the above experiments by using the same format for each query. We first provide the question description directly. In addition to the question requirements, the description may also contain some hints to students. We do not have to specify that we use and want answers in OCaml, since ChatGPT is able to identify the programming language from the given starter code or problem description. The following shows an example request to ChatGPT for the \texttt{Binomial} problem.




\begin{figure}[h]
    \centering
    \includegraphics[scale=0.83, trim={0.2cm 21.5cm 10.8cm 0cm}, clip]{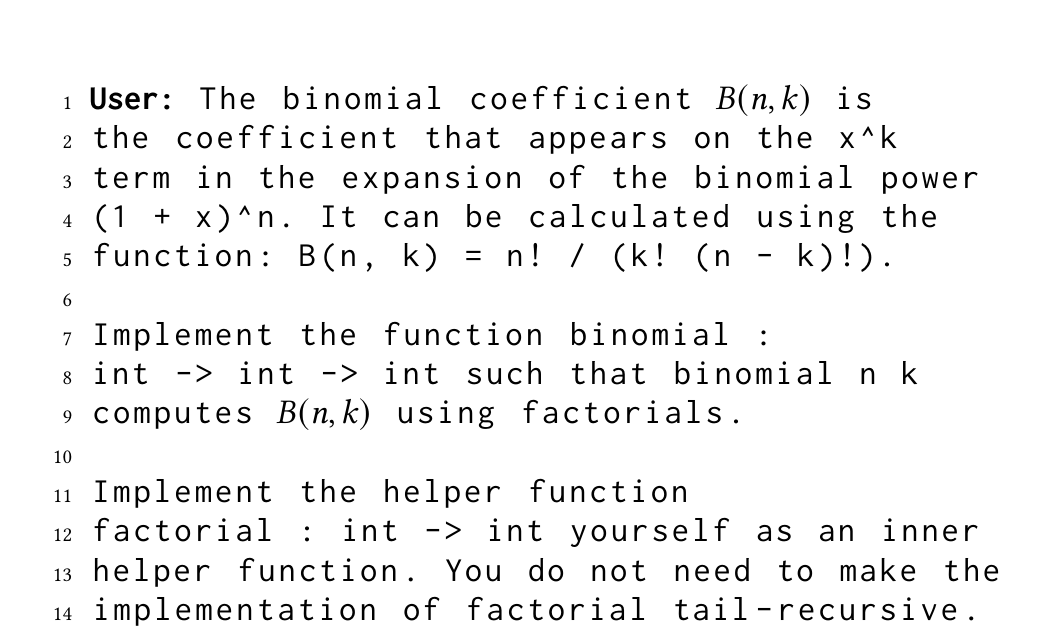}
    \caption{Request to ChatGPT for the Binomial problem from 2022 Fall homework.}
    \label{fig:binomial}
\end{figure}

In the second phase of our work we evaluate solutions automatically produced by using \emph{"assisted"} ChatGPT. Inspired by prompt engineering, here we guide ChatGPT towards an answer, similarly to a TA guiding a student during office hours. This evaluates the potential of ChatGPT as a learning-assisted tool, i.e. how interacting with it improves its final answer. We adopt various strategies to improve the model's response given the diversity of question types in this course. Those methods can be summarized as \emph{paraphrasing input question}, \emph{providing natural language hint}, \emph{teaching by examples} and \emph{providing test cases}. We believe that these strategies are effective because paraphrasing enables more precise comprehension of input questions by ChatGPT. Test cases are particularly useful for programming tasks, allowing the model to analyze both the logic and function signature in complex types. Teaching by example provides more structure. In particular for induction proof it aids the model in learning correct base cases and step cases. Finally, providing natural language hints, such as pointing out type mismatch in the solution -- frequently done also by instructors and TAs when interacting with students -- 
is also a technique for improving the model's solutions. 

\section{Unassisted Experiment}







\begin{table*}[h]
\begin{tabular}{|l|l|l|c|l|l|l|}
 \hline
 \multicolumn{7}{|c|}{\textbf{Homework Results from ChatGPT (Unassisted Mode)}} \\
 \hline
 HW & Topic &Question Description &Score (/100\%) & E1 & E2 & E3\\
 \hline
 
 \multirow{3}{*}{1}& \multirow{3}{*}{Basics}&Fix two incorrectly implemented functions &  100.0 & 0 & 0 & 0\\
 & &Compute binomial coefficients&  100.0 & 0 & 0 & 0\\
 & &Compute Lucas numbers (tail-recursive) &  92.6 & 0 & 0 & 1\\
 \cline{1-7}
 
 \multirow{4}{*}{2}& \multirow{4}{*}{Pattern Matching \& Recursion}&convert between unary numbers and integers & 100.0 & 0 & 0 & 0\\
 &   &Manipulate mathematical expressions&  100.0 & 0 & 0 & 0\\
 &  &Evaluate mathematical expressions&  100.0 & 0 & 0 & 0\\
 &  &Compute the derivative of expressions &  86.7 & 0 & 0 & 1\\
\cline{1-7}

 \multirow{2}{*}{3}& \multirow{2}{*}{Church Numerals}&Manipulate with church numerals and integers& 0.0 & 0 & 1&  0\\
 &  &Sum a list of church numerals & 0.0 & 0 & 1&  0\\
\cline{1-7}
 
 \multirow{3}{*}{4}& \multirow{3}{*}{Backtracking}&Obtain variable names in the formula&  100.0 & 0 & 0 & 0\\
  &   &Evaluate Boolean formulae& 93.3 & 0 & 0 & 1\\
  &   &Identify true assignments&  0.0 & 0 & 1 & 0\\
\cline{1-7}
 
 \multirow{3}{*}{5}& \multirow{3}{*}{Continuation}& Find the depth of binary trees& 100.0 & 0 & 0 & 0\\
 &   & Find the number of sub-trees& 100.0 & 0 & 0 & 0\\
 &   &Implement a simple parser&  0.0 & 0 & 0 & 5\\
\cline{1-7}
 
 \multirow{4}{*}{6}& \multirow{4}{*}{Data Transformations} & Generalization of map to transform a list&  100.0 & 0 & 0 &0\\
 &   &Manipulate with Maybe.t values&  80.0 & 0 & 1 &0\\
 &   &Explore some explicit isomorphisms & 66.7 & 0 & 2 &0\\
 &   &More practice with continuations & 100.0 & 0 & 0 &0\\
\cline{1-7}
 
  \multirow{2}{*}{7}& \multirow{2}{*}{HOF \& Reference} &Conversion between string and characters& 100.0 & 0 & 0 &0\\
  &  &Create a bank account  & 100.0 & 0 & 0 & 0\\
 \cline{1-7}
    
 \multirow{4}{*}{8}& \multirow{4}{*}{Lazy programming \& Streams  }&Peek/drop the 1st element in a list&  100.0 & 0 & 0 & 0\\
 & &Implement number streams& 100.0 & 0 & 0 & 0\\
 & &Map two streams to a transformed stream& 100.0 & 0 & 0 & 0\\
 & &Implement Fibonacci and Lucas number streams& 100.0 & 0 & 0 & 0\\
\cline{1-7}

 \multirow{3}{*}{9}& \multirow{3}{*}{MiniCaml part 1}&Collect free variable& 95.0 & 0 &0  & 1\\
  &   &Collect unused variable& 0.0 & 0 &1  &0\\
  &   &Perform substitution& 0.0 & 0 &1  &0\\
\cline{1-7}
  
  \multirow{3}{*}{10}& \multirow{3}{*}{MiniCaml part 2}&Evaluate values based on big-step evaluation& 0.0 & 0 &1  &0\\
  &   &Infer the type of an expression& 0.0 & 1 &0  &0\\
  &   &Unification& 0.0 & 0 &1  &0\\
\cline{1-7}
  
 \hline
 \multicolumn{7}{|c|}{ \# HW with 100\% = 16 (out of 31)} \\
 \hline
\end{tabular}
\caption{\label{hw-table}: HW results from ChatGPT. Here we use some abbreviations: E1=syntax error, E2=compilation error, and E3=logical error. The first column represents the number of the homework assignment from which the question is taken. The second column indicates the material covered in this homework while the third provides brief descriptions of each task. The last four columns record the total score the model achieves, as well as the number of syntax errors, compilation errors and logical errors respectively.}
\end{table*}

In this section, we report our results for unassisted experiments.

Table \ref{hw-table} shows the grades obtained by ChatGPT for each homework questions and the corresponding errors in its solutions. Each of the question is auto-graded by Learn-OCaml platform and the point is then calculated
out of 100\%. The last 3 column indicates the number of errors in each error category which are syntax error, compilation error and logical
error. As we can see, ChatGPT is able to give the perfect solution for more than half (16 out of 31) the problems. And for another 4 tasks it obtains 85\% or higher. We also notice that the model obtains 0 for 9
question mainly due to compilation errors which means the solution cannot be graded by the auto-grader. In particular, the model performs the worst on homework 3, 9 and 10. Among these three assignments, homework 3 is quite a special case because we define \emph{Church numerals} in a different way from to the most common formulations. In particular, we swap the two input arguments to the function that defines a Church numeral. However, the model seems to ignore the definition given but interprets \emph{Church numerals} using the conventional definition. As a result, its produced solution fails to
type-check in homework 3. As for homework 9 and 10, they involve
implementing a programming language called MiniCAML. These homework
questions rely on understanding type checking and type inference
concepts. This is in contrast, in homework 7 and 8, which cover HOF, reference, lazy programming and streams. While these topics are usually considered challenging for students, ChatGPT successfully passes all the tests. Through what the model achieves above, we can conclude that ChatGPT seems to be good at figuring out complex logic behind functions, and generate neat and elegant code. However, it struggles at deducing types of expressions.


Overall, ChatGPT was very effective in generating recursive functions
for basic problems involving pattern matching; it vastly exceeded our
expectations on topics like tail-recursion, use of HOF, lazy
programming, and even continuations which are usually considered very challenging for
students. This was indeed very surprising to us. ChatGPT provides in
fact elegant solutions to short complex problems that are usually
non-trivial and difficult for students. The generated solutions also
often were breaking up a complex problem into smaller helper
functions, and ChatGPT had no difficulty to use state (references) and
exceptions for backtracking problems. It was also surprising to us
that most of the time when ChatGPT struggled it was due to type
violations in the generated code. This however seems potentially easily
addressed by providing assistance (see Section \ref{sec:ass}) or validating the
answers and incorporating a form of formal reasoning into natural
language models \cite{sap2020commonsense,han2022folio}.



Table \ref{tab:exam-table} presents what ChatGPT achieves on both
midterm and final exam in 2022 Fall term in unassisted experiment. For
each exam, questions are categorized into 3 types: conceptual
questions (multiple choice or short answer), code completion questions
and  proof (shown in column 2). In column 3, we give a brief
description for each question, following by the score obtained in
column 4. From the table we can see that the model's performance is
impressive for code completion part of both exams, but it struggles
when it comes to conceptual questions. One reason could be that coding
questions on exams are likely to have shorter and precise
instructions.
Conversely, questions on OCaml concepts require a deeper understanding of OCaml paradigms. These findings on exam are consistent with those on homework.

\begin{table*}[h]
    \centering
    \large
\begin{tabular}{ |l|l|l|l| } 
\hline
\multicolumn{4}{|c|}{\textbf{Exam Results from ChatGPT (Unassisted Mode)}} \\
\hline
Exam &Qs Type &Question & Score \\
\hline
\multirow{10}{*}{Final Exam} 
& \multirow{4}{*}{Conceptual Qs} & Identification of types in expressions & 7.5/13 \\ 
& &Result of function bindings & 4.0/12\\ 
& & Identification of subtypes & 4.0/12 \\ 
& & Substitution & 0.0/8\\ \cline{2-4}

& \multicolumn{1}{l|}{\multirow{4}{*}{Code Completion}} &Higher-order functions & 35.0/40 \\ 
& \multicolumn{1}{c|}{} & Object-oriented programming&  13.0/20 \\ 
& \multicolumn{1}{c|}{} & Memorization & 15.0/15 \\ \cline{2-4}

& \multirow{1}{*}{Induction proof} & Induction on mobile & 15/20 \\
\hline

\multirow{4}{4em}{Midterm}
& \multirow{2}{*}{Conceptual Qs} 
& Identification of types in expressions & 12.5/20 \\
& &Short answer questions on theory& 7.0/20\\ \cline{2-4}

& \multirow{1}{*}{Code Completion} & Storing a set of strings using continuation & 15.0/20 \\ \cline{2-4}

& \multirow{1}{*}{Induction proof} & Induction on higher-order function map & 15.0/20 \\
\hline
\end{tabular}
    \caption{Overview of what ChatGPT achieves on final exam and
      midterm grouped by question (Qs) type.
 }
    \label{tab:exam-table}
\end{table*}

\begin{figure}[h]
    \centering
    \includegraphics[scale=0.65,trim={1cm 12cm 9cm 0.8cm}, clip]{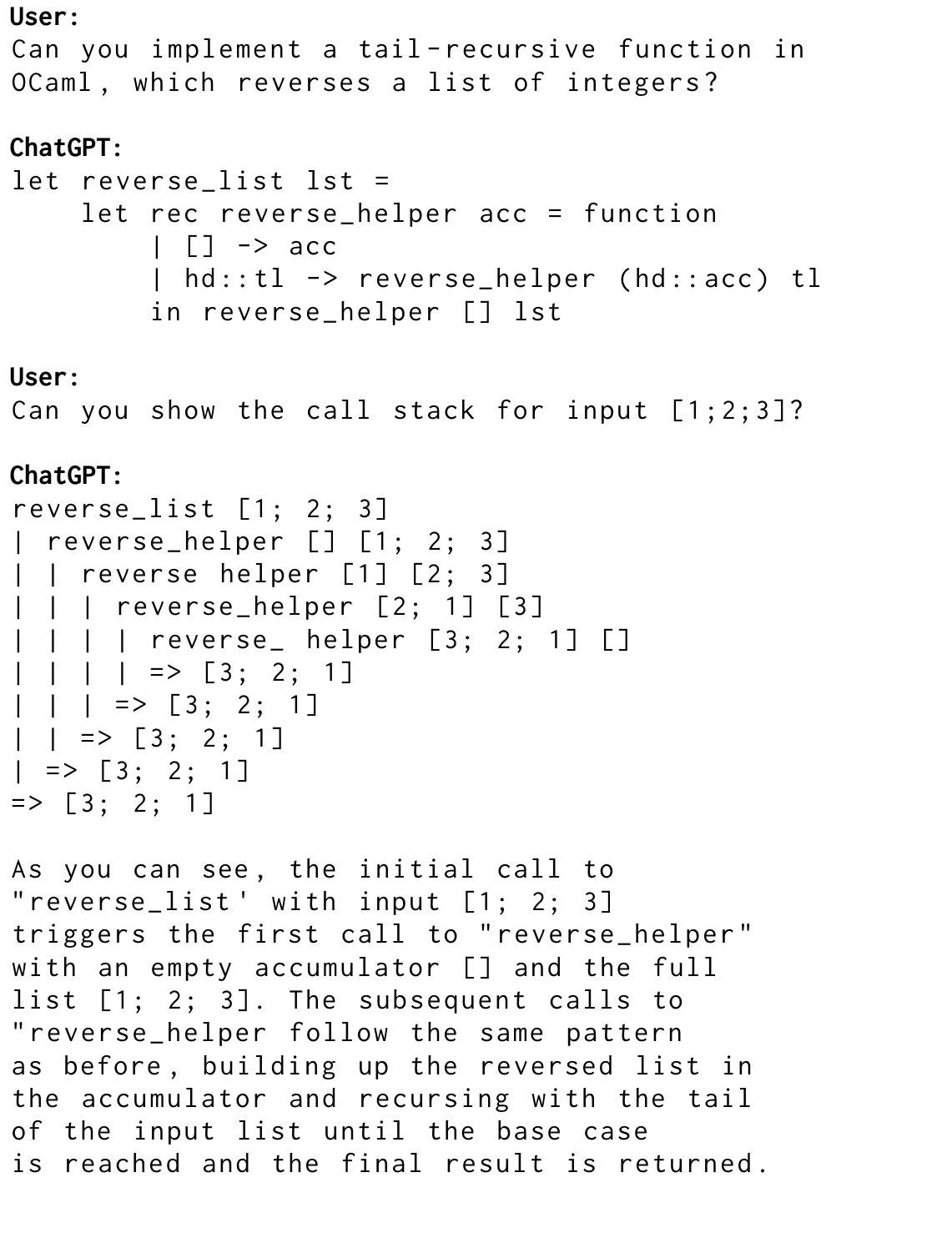}
    \caption{A request for ChatGPT to generate the call stack for a tail-recursive function.}
    \label{fig:cs1}
\end{figure}

Provided that ChatGPT is highly effective at coding tasks on exams and homework, this raises two natural questions: 1) Does the model really understand the code it generates? 2) Can it provide explanations of how a program is executed? To get an impression, we asks ChatGPT to print out the call stacks as shown in Figure \ref{fig:cs1}. Here ChatGPT produces the calls of functions for a tail-recursive function it implements, followed by the corresponding return. The result is impressive, as it correctly unfolds the recursive call and generates an execution trace similar to what a TA might draw on the board to explain the execution of a recursive program. As a result, the model demonstrates a degree of understanding of the function it produces. However, caution is necessary as the call stack it outputs is not always reliable.

\section{Assisted Experiments}
\label{sec:ass}
\subsection{Assisted Experiment Methods}

Prompt learning and engineering have proven to be effectively useful
and became a major area in NLP/LLM research
\cite{reynolds2021prompt}. In other words, the sole API between users
and models is only made by a series of (text) inputs which largely
affect models' performance. In this paper, we also want to demonstrate
the importance of prompt engineering by showing that ChatGPT can 
improve its answers if more relevant information is given. In our
so-called assisted study, we develop four methods that are
paraphrasing, providing natural language hint, teaching by example and
providing test cases. Using these techniques, we re-evaluate ChatGPT
on all the cases where the answer previously generated by it did not achieve a perfect score. 
%

\begin{figure*}[h]
  \centering
  \begin{minipage}[h]{0.48\textwidth}
    \centering
    \includegraphics[width=\textwidth, trim={0 19cm 9cm 0.7cm}, clip]{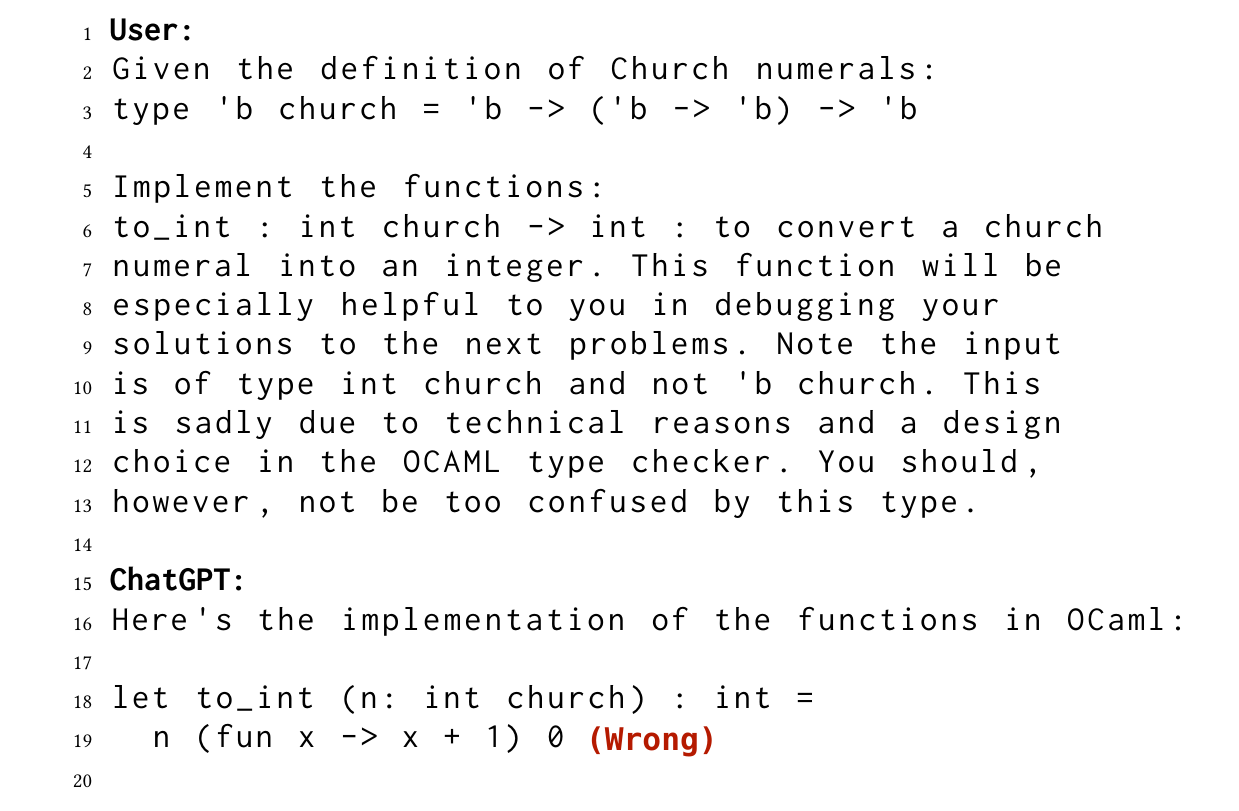}
    \subcaption{Church numeral question (original).}
  \end{minipage}
  \hfill
  \begin{minipage}[h]{0.48\textwidth}
    \centering
    \includegraphics[width=\textwidth, trim={0 10.8cm 9cm 9.2cm}, clip]{figures/para_full.pdf}
    \subcaption{Charles numeral question (after paraphrasing).}
  \end{minipage}
  \caption{An example of paraphrasing a programming question from HW3.}
  \label{fig:para_full}
\end{figure*}

We find that paraphrasing requests can help ChatGPT provide more accurate responses. Figure \ref{fig:para_full} illustrates a conversation with ChatGPT about a programming question before and after paraphrasing. In this case, the problem was unique because we defined Church numerals in a different way than usual. As a result, ChatGPT's initial response (lines 18) was incorrect as it employed the common definition of Church numerals in its answer. However, after we changed the name to \emph{Charles} instead of \emph{Church}, the model correctly interpret our new definition and provide the correct answer to the question (line 39).

\begin{figure}[h]
    \centering
\includegraphics[width=0.52\textwidth,trim={0.5cm 12cm 9cm 0.3cm},clip]{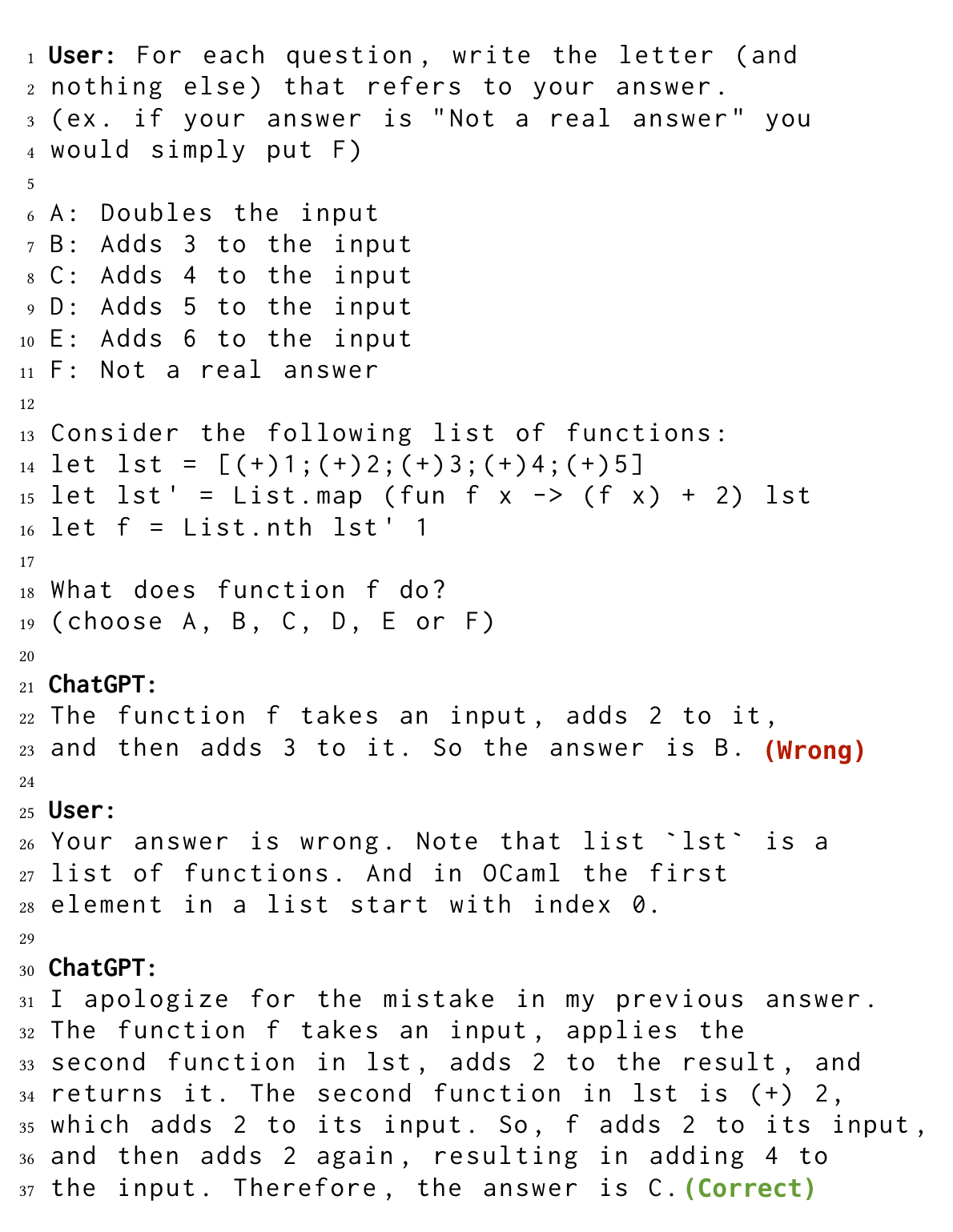}
    \caption{An example of providing natural language hint to a MCQ from midterm exam.}
    \label{fig:hint_full}
\end{figure}

Providing natural language hint is the most commonly-used way in instructors' office hours to guide students towards the correct answer. Figure \ref{fig:hint_full} illustrates the interaction with ChatGPT of a MCQ from the midterm (lines 1-19). In the first response (lines 22-23), ChatGPT fails to interpret what function \texttt{f} does or what variable \texttt{lst} consists of. We give ChatGPT a hint of what the list \texttt{lst} contains and provide some basics for lists in OCaml (25-28), which helps the model break the \texttt{f} function into smaller steps, resulting in a correct response returned by the model (lines 30-37) with clear explanation.

\begin{figure*}[h]
  \centering
  \begin{minipage}[h]{0.48\textwidth}
    \centering
    \includegraphics[width=1.05\textwidth, trim={0 4cm 9.3cm 0cm}, clip]{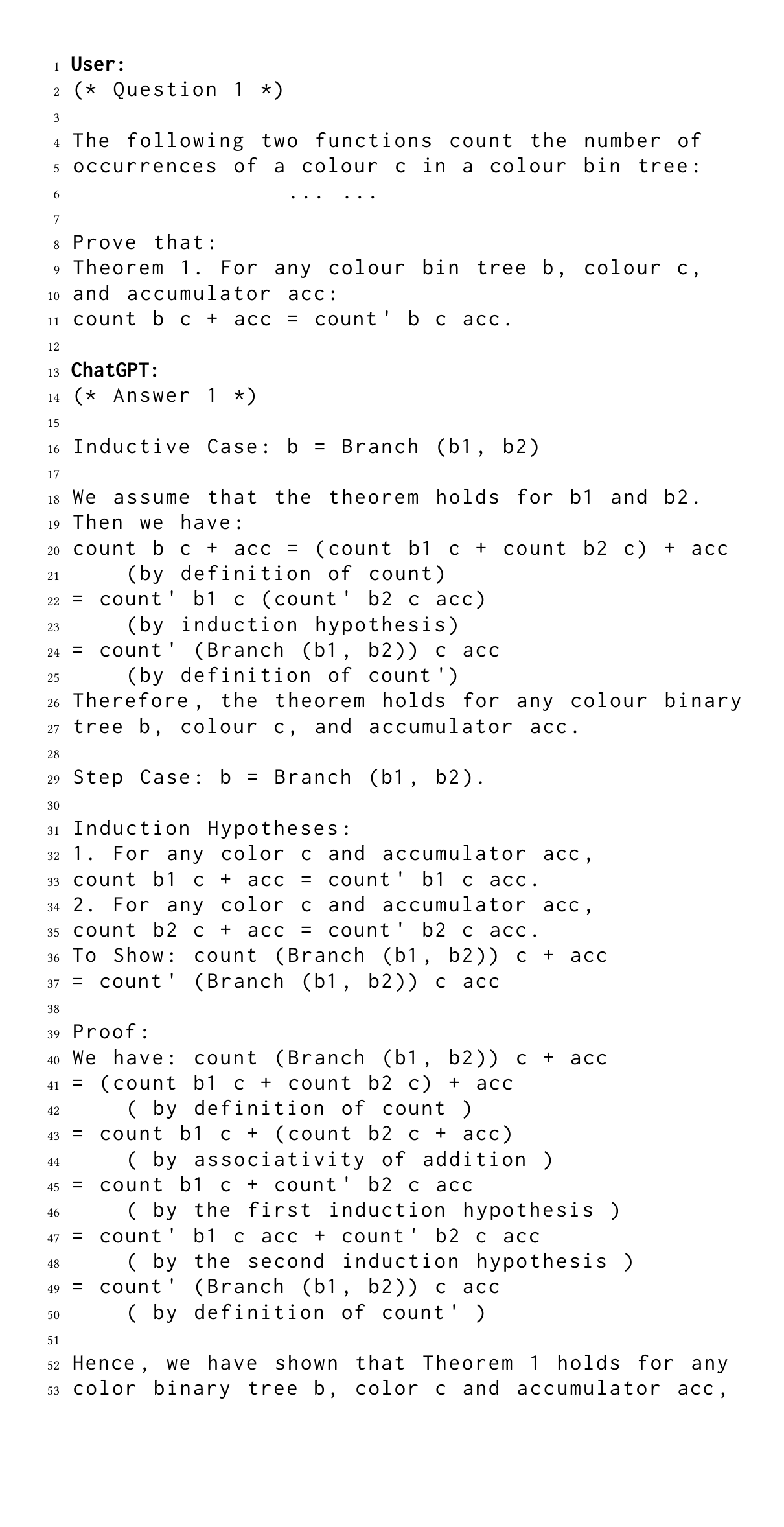}
    \subcaption{Induction proof before providing exemplar.}
    \label{fig:ind_a}
  \end{minipage}
  \hfill
  \begin{minipage}[h]{0.48\textwidth}
    \centering
    \includegraphics[width=1.05\textwidth, trim={0.8 2cm 8cm 0}, clip]{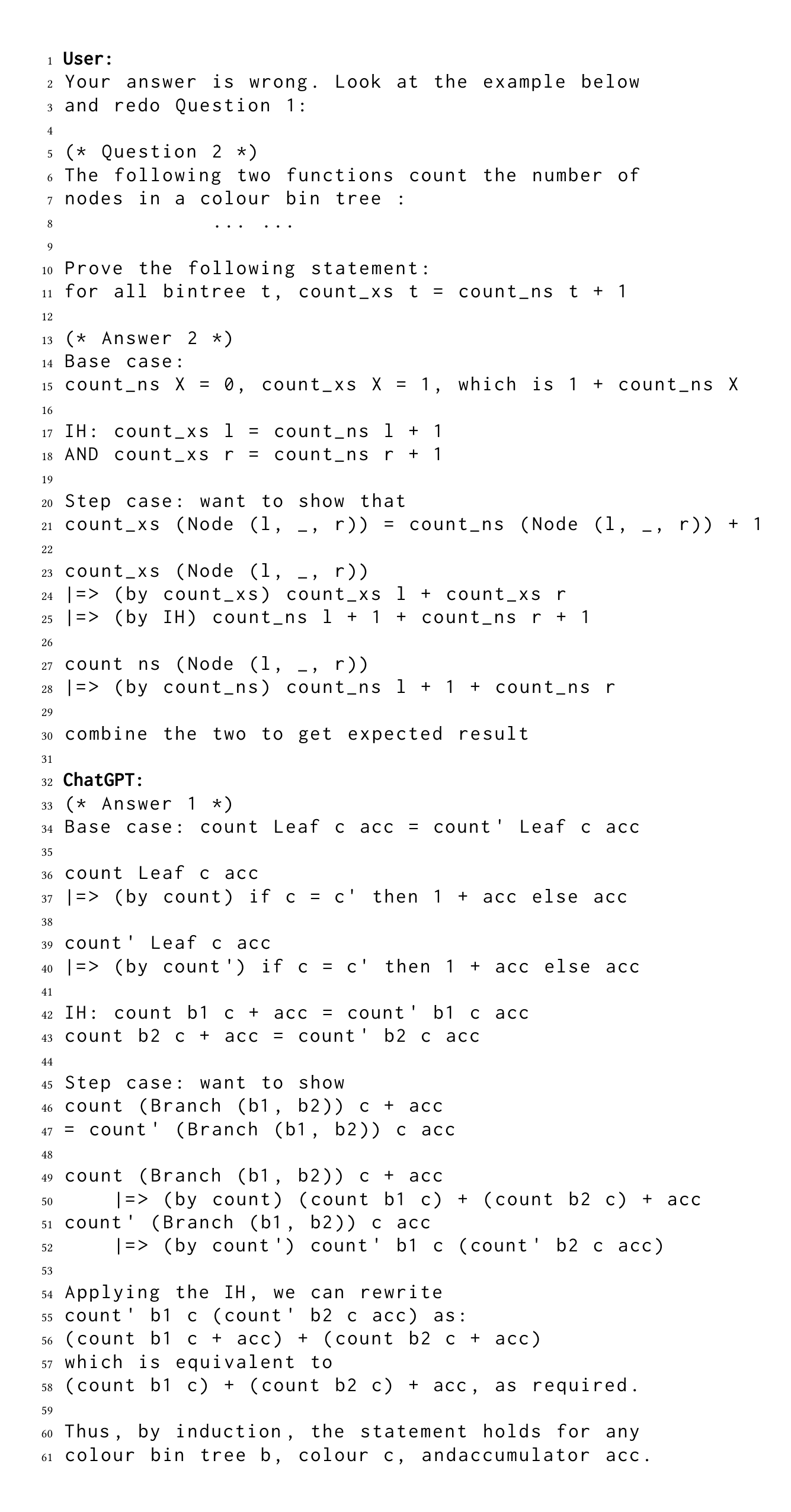}
    \subcaption{Induction proof after providing exemplar.}
    \label{fig:ind_b}
  \end{minipage}
  \caption{An illustration of teaching by example of an induction proof from the final exam.}
  \label{fig:ind_full}
\end{figure*}

In order to teach the model by example, it will be given a in-context exemplar of input–output pair before outputting a prediction for a test-time example. Exemplar is formatted as question and answer. In our case, we select an example induction question from the course tutorial. Figure \ref{fig:ind_a} presents the original solution returned by ChatGPT. Here the model fails to identify the correct base case and the response is not consistent with the standard structure of an inductive proof. We then input the exemplar which is also an induction proof problem on counting the number of leaves and internal nodes respectively (as shown in line 1-30 in Figure \ref{fig:ind_b}). After providing this example the new solution returned (line 33-63) adheres to the typical form of such proofs, and the proof consists of logical reasoning and clear deductions, which finally reaches to the required conclusion.

In addition to the three methods shown above, we mainly provide test cases on the programming tasks from homework, which will be presented in the next section.

\subsection{Assisted Experiment Results}

\begin{table*}[h!]
 
\begin{tabular}{ |c|l|l|c|c|c| }
 \hline
 \multicolumn{6}{|c|}{\textbf{Homework Results from ChatGPT (Assisted Mode)}} \\
 \hline
 
 HW & Problem description& Improved Score & Paraphrasing &Providing Hint& Test Cases \\
 \hline
 
 1 & Compute Lucas numbers(tail-recursive)
 & 100.0 (+7.4)& &  &\checkmark\\
 \hline
 2 & Compute the derivative of expressions
 & 100.0 (+13.3)& &  &\checkmark\\
 \hline

 \multirow{2}{*}{3}
 & Manipulate with church numerals and integers
 &66.7 (+66.7)&\checkmark &  &\\ 
 &Sum a list of church numerals
 &100.0 (+100.0) &\checkmark& & \checkmark\\ \hline
 
 \multirow{2}{*}{4}
 &Evaluate Boolean formulae
 & 100.0 (+6.7)&  & &\checkmark\\ 
 &Identify true assignments
 &100.0 (+100.0)& &\checkmark &\checkmark\\ \hline
 
 5 & Implement a simple parser
 &0.0 (-) & - & - & -\\
 \hline

 \multirow{2}{*}{6}
 & Manipulate with Maybe.t values
 &100.0 (+20.0)&\checkmark &\checkmark &\\ \cline{2-6}
 &Explore some explicit isomorphism
 &100.0 (+32.3)& &\checkmark &\\ \hline
 
 9 & MiniCaml part 1&33.3 (-)& - & - & - \\
 \hline
 10 & MiniCaml part 2 &0.0 (-)& - & - & -\\
 \hline
\end{tabular}
\caption{\label{prompt_hw}Overview of prompt learning results for homework tasks. Here we use \checkmark to indicate the methods used for each question for improving ChatGPT's solutions. Note that we may need to use more than one method for each task. If there is no \checkmark in a row that means the solution for this task fails to be enhanced by any of the four methods in the table.}
\end{table*}

Table \ref{prompt_hw} shows the results of using four methods to improve the performance of ChatGPT on different homework tasks. The table includes information about the homework problem, the initial score, and the score after the methods are applied. The table also indicates which methods are used for each question. It can be noticed that some topics are easier for ChatGPT to improve than others. For example, HW3 and HW4 have perfect scores after the methods are applied, indicating that the methods are effective in improving ChatGPT's solutions. In contrast, HW5 has a score of 0 even after applying the methods, suggesting that ChatGPT struggles to find a solution continuations. HW9 also has a relatively low score after the methods are applied, suggesting that designing MiniCaml is a difficult problem for ChatGPT to solve.

In terms of the methods used, it seems that providing hints and manipulating the input question are the most effective techniques. These methods are used for most of the questions in the table, and they are often used in combination with other methods. Paraphrasing are also used for several questions, but it is must be used along with other methods. Test cases are also proved to be effective as HW1-4 are significantly improved by this strategy.

\begin{table*}[h!]
    \centering
    \begin{adjustbox}{width=0.97\textwidth}
\begin{tabular}{ |c|l|l|c|c|c| } 
\hline
\multicolumn{6}{|c|}{\textbf{Exam Results from ChatGPT (Assisted Mode)} }\\ \hline
\hline
\multicolumn{6}{|c|}{\textbf{Final Exam}}\\
\hline 
Qs Type & Question&Improved Score& Paraphrasing& Providing Hint& Teaching by Example\\
\hline

\multirow{4}{*}{Conceptual Qs}
& Identification of types in expressions & 7.5/13 (-) & - & - & - \\ 
& Result of function bindings & 8.0/12 (+4.0)
&\checkmark &\checkmark &\\ 
& Identification of subtypes & 4.0/12 (-)
& - & - & -\\ 
& Substitution & 8.0/8 (+8.0)
& & &\checkmark\\ \cline{1-6}

\multirow{2}{*}{Code Completion}
& Higher-order functions & 40.0/40 (+5.0)
& &\checkmark &\\ 
& Object-oriented programming & 15.0/20 (+2.0)
&\checkmark & \checkmark&\\ \cline{1-6}

\multirow{1}{*}{Proof}
& Induction on mobile & 20.0/20 (+5.0)
& & &\checkmark\\ 
\hline
\hline

\multicolumn{6}{|c|}{\textbf{Midterm}}\\ \hline
\multirow{2}{*}{Conceptual Qs}
& Identification of types in expressions & 15.0/20 (+2.5) & \checkmark& \checkmark& \\ 
& Short answer questions on theory & 10.0/20 (+3.0)
&\checkmark &\checkmark &\\ \hline

\end{tabular}
\end{adjustbox}
    \caption{Results of exam questions in assisted mode. We use \checkmark to indicate the method applied for improvement for each question. If all the four methods fail to assist the model providing better solutions we simply put - in this line. Among these three methods, Providing Hint is the most frequently used as it is checked for almost all the question types. }
    \label{tab:exam_after}
\end{table*}

Table \ref{tab:exam_after} presents the result of assisted experiments for midterm and final exam. From the table, we can observe that the methods used for improving the grades of exams include Paraphrasing, Providing Hint, and Teaching by Example. Among these three methods, Providing Hint is the most frequently used, as it is checked for almost all questions in the final exam section and for two questions in the midterm section. Paraphrasing is used for only two questions in the final exam section, while Teaching by Example is used for two questions in the final exam section and one question in the midterm section. Overall, final exam score is significantly improved in assisted study while the midterm is also improved to some extent.

\subsection{Improvement}

\begin{table*}[h!]
    \centering
    \begin{tabular}{|c|l|l|l|l|l|l|}
      \hline
     \multicolumn{7}{|c|}{\textbf{Comparison of Exam Results in Unassisted/Assisted Experiments}} \\
     \hline
     Exam & Mode &Total Score (/100\%)& Total Score& Conceptual Qs & Code Completion & Proof\\
     \hline
     \multirow{2}{4em}{Final Exam}
      & Unassisted & 64.4  &93.5/140 &  15.5/45 & 63.5/75 & 15.0/20\\
      & Assisted &81.0 (+16.6) &117.5/145 (+22.0) &  27.5/45 (+12.0) & 70.5/75 (+7.0)& 20.0/20 (+5.0)\\
     \hline
     \multirow{2}{4em}{Midterm}
     & Unassisted & 74.5& 74.5&19.5/40& 30.0/30 & 25.0/25\\
     & Assisted & 79.5 (+5.0)& 79.5 (+5.0)&25/40 (+5.5)& 30.0/30 (-)& 25.0/25 (-)\\
     \hline
    \end{tabular}
    \caption{Comparison of exam statistics in unassisted/assisted experiments. Here we indicate the amount increased in brackets for the assisted study results, and (-) simply means there is no change in score.}
    \label{tab:exam_compare}
\end{table*}

Table \ref{tab:exam_compare} compares the results from two different studies - one unassisted and the other assisted. The results are based on two different exams, the final exam and the midterm exam.

For the final exam, the assisted result outperforms the unassisted result by a significant margin. The assisted group's total score is 81.0, which is 16.6 points higher than the unassisted group's score of 64.4. The assisted group also performs better on all three types of questions - conceptual, code completion, and induction-proof. The conceptual score is 27.5 for the assisted group, which is 12.0 points higher than the unassisted group. The code completion score is 70.5 for the assisted group, which is 7.0 points higher than the unassisted group. Finally, the induction proof score is 20.0 for the assisted group, which is 5.0 points higher than the unassisted group. For the midterm exam, the assisted group also outperforms the unassisted group, but the difference is not as significant. The assisted group's total score is 79.5, which is 5.0 points higher than the unassisted group's score of 74.5. The assisted group performs better on the conceptual questions, with a score of 25.0, which is 5.5 points higher than the unassisted group. 

\begin{table}[h!]
    \centering
    \begin{adjustbox}{width=0.5\textwidth}
    \begin{tabular}{|c|c|c|c|c|}
    \hline
     Mode  & Homework & Midterm & Final &Total Grade \\
    \hline
    Unassisted &271/314& 225/314& 170/314 & 220/314 \\
    \hline
    Assisted & 233/314& 155/314 & 121/314 & 155/314 \\
    \hline
    \end{tabular}
    \end{adjustbox}
    \caption{Rank of ChatGPT in unassisted/assisted mode.}
    \label{tab:compare_rank}
\end{table}

Figure \ref{tab:compare_rank} compares the rank of homework, midterm and final exam achieved by ChatGPT in both modes. In the unassisted mode, ChatGPT ranks lower in all four evaluation criteria compared to the assisted mode. The biggest difference in rank can be observed in the final exam, where ChatGPT ranks 170th in the unassisted mode and 121th in the assisted mode. This suggests that the methods used in the assisted mode have improved ChatGPT's performance in the final exam. Similarly, in the assisted mode, ChatGPT ranks higher in all four evaluation criteria compared to the unassisted mode. As for the total grade, ChatGPT ranks 155th in the assisted mode and 220th in the unassisted mode. This indicates that the methods used in the assisted mode help ChatGPT to perform better overall.

\begin{figure*}[h!]
    \centering
    \includegraphics[scale=0.4, trim={1cm 0cm 0.8cm 1.82cm},clip]{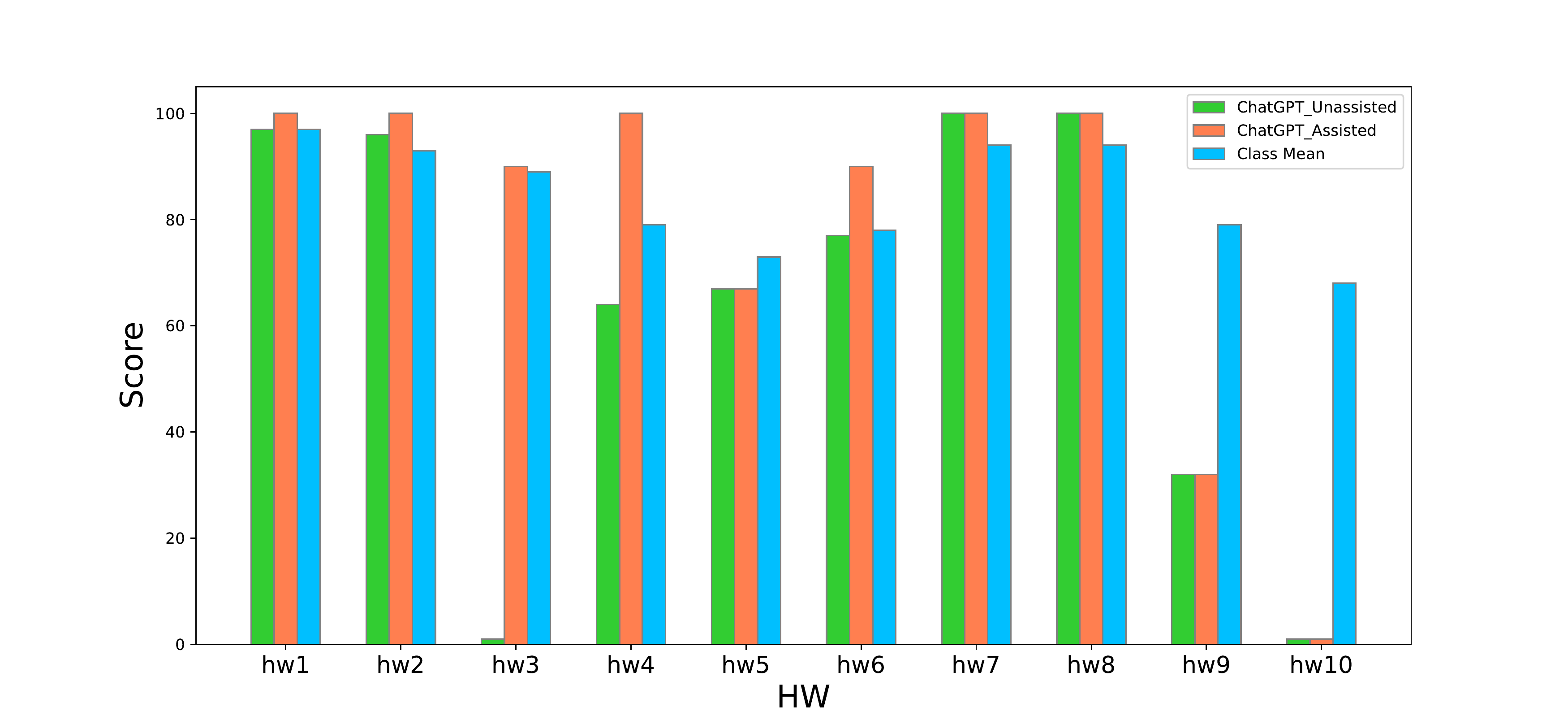}
    \caption{Grade comparison for homework of ChatGPT unassisted/assisted with the class mean.}
    \label{fig:compare_hw}
\end{figure*}

As for homework, the overall rank does not give us much information as the performance of the model deviates a lot across ten homework. Figure \ref{fig:compare_hw} is a bar plot that compares the grades of ChatGPT in two different studies (ChatGPT unassisted and ChatGPT assisted) with the class mean. The heights of the bars represent the scores of the model and the class mean on 10 different homework assignments (HW1 to HW10). The plot provided shows that the model performed significantly better in the assisted study compared to the unassisted study.

Comparing the two sets of data to the class mean, we see that the model achieve grades higher than the class mean for most of the homework assignments in both studies. However, the improvement in the assisted mode is even more noticeable, with the model achieving grades equal to or higher than the class mean for all of the homework assignments in the assisted study, while it only achieves this for 7 out of 10 homework assignments in the unassisted study.


\section{How Can ChatGPT Improve Students' Learning Experience ?}

With its ability to provide personalized support and feedback on assignments and exams, ChatGPT has the potential to enhance students' understanding of OCaml programming concepts and improve their overall performance in the course. We summarize the following aspects in which ChatGPT could assist students' learning:

\textbf{Offering real-time support:} As a language model, ChatGPT can provide instantaneous help to students by fixing errors in their code. For example, if a student is working with a programming language that they are not familiar with, ChatGPT can provide assistance by clarifying error messages and offering suggestions for how to fix them. ChatGPT can also provide syntax tips to help students write correct and efficient code. This includes suggestions for common programming structures and best practices, as well as guidance on how to avoid syntax errors. 
In general having access to immediate feedback on their code has been recognized to significantly improve student learning outcomes and engagement with their assignments (see, e.g., \cite{sherman2013impact, wilcox15_role_autom_under_comput_scien_educat, Gramoli_2016}). In addition, ChatGPT can suggest alternative solutions to programming problems, which can be particularly helpful for students who are struggling with complex programs or need help debugging their code.


 
 Overall, this real-time support can be particularly helpful for students who are struggling with complex programs or need help debugging their code. This support can prove valuable in keeping students focused, tackling obstacles, and receiving timely feedback on their progress \cite{fernoagua2018intelligent}. For students who lack access to immediate assistance from instructors or teaching assistants, ChatGPT can be an indispensable resource. By addressing students' programming-related questions, ChatGPT can provide comprehensive explanations and examples to help them grasp programming concepts more effectively. 
 



\textbf{Solving problems step by step:} Learning programming can be challenging, especially when dealing with complex problems. Fortunately, ChatGPT can help by breaking down these problems into smaller, more manageable parts using Chain-of-Thought prompting \cite{wei2022chain}. This technique encourages step-by-step thinking and allows ChatGPT to provide users with step-by-step solutions that help them work through the problem and arrive at a solution. For instance, if a user inputs a complex programming problem, ChatGPT can analyze the problem and generate a response that outlines the steps the user should take to solve it. As the user inputs feedback and follow-up questions, ChatGPT can generate additional steps that help them work through the problem in a logical and efficient way. By breaking down problems into smaller parts and generating step-by-step solutions, ChatGPT can help users better understand how to approach programming tasks.

In addition to the above-mentioned aspects, we also see strong potential in the following two directions which would be useful to explore in more depth in the future. Firstly, ChatGPT can be utilized to generate exercises that are customized to a student's individual skill level, allowing them to practice and apply programming concepts in a targeted and effective manner. By offering structured and meaningful exercises, ChatGPT can assist students in building their programming skills in a more focused and efficient way. Secondly, ChatGPT can provide personalized feedback and guidance to students based on their unique needs, enhancing their learning experience. By allowing students to input their own prompts and questions, ChatGPT can help them learn at their own pace and in a manner that is tailored to their learning style.

Overall, ChatGPT can be a valuable learning assistant for students learning to code. With its help, students can gain a deeper understanding of programming concepts and develop the skills needed to succeed in their coursework and future careers.

\section{How Can Instructors
Take Advantage Of 
ChatGPT?}


Like students, ChatGPT can be an invaluable tool for instructors and teaching assistants in many ways. However, due to the unreliability of ChatGPT, some careful examinations are needed before instructors can take advantage of it. Thus,
we only highlight some potential areas of its usage.


\textbf{Creating exercises/homework/assignments:} 
This feature of ChatGPT can save teachers time and effort in creating assignments and assessments for their students. By simply inputting a prompt or a topic, ChatGPT can generate a range of questions with varying levels of difficulty and complexity. Given the fact the generated exercises or homework may be incorrect, instructors and TAs can manually go over its generation and choose those that are correct and appropriate for their students, providing them with a challenging and engaging learning experience.




\textbf{Answering student questions:} By using ChatGPT to answer student questions, teachers can save time and focus on other tasks while still ensuring that students receive timely and accurate feedback. This is especially helpful for larger classes such as massive open online courses (MOOCs) where teachers may not be able to provide individual attention to each student \cite{Geng_2023}. 


\textbf{Generating lesson plans:} Using ChatGPT to generate lesson plans can be a time-saving and helpful tool for teachers. With its vast knowledge base, ChatGPT can provide a variety of suggestions for activities, materials, and resources that teachers can incorporate into their lesson plans. This can ensure that teachers are providing comprehensive instruction and engaging students in different ways.

Overall, ChatGPT can be a valuable tool for teachers, providing a range of benefits that can help save time, improve instruction, and personalize learning for individual students.

\section{Discussion}
There are currently two extreme views on the potential of ChatGPT. One perspective sees ChatGPT as a stochastic parrot that merely remembers the probability distribution of various contexts \cite{DBLP:conf/fat/BenderGMS21,aboufoul_2022} while the other perceives it as a possible replacement for coders and a solution to all problems \cite{mok_2023}. The latter perspective naturally raises the question of whether programming education is still essential in light of advanced language models like ChatGPT.

\begin{figure}[h]
  \centering
    \includegraphics[width=0.45\textwidth]{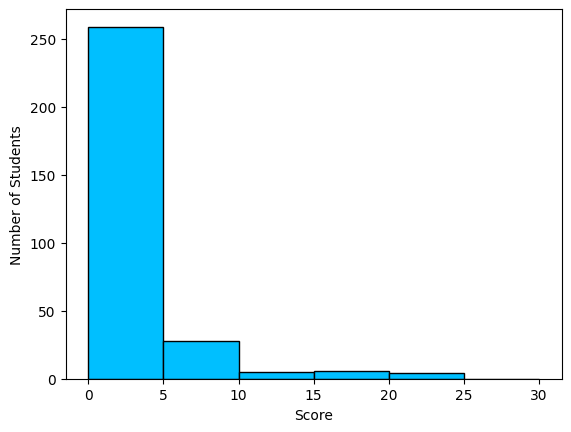}
    \caption{The distribution of HW grade improvement assuming students have access to ChatGPT.}
    \label{fig:imp_hw}
\end{figure}

However, as illustrated in Figure \ref{fig:imp_hw}, using ChatGPT did not result in a significant improvement in students' homework grades as one might have expected.  Moreover, upon examining Table \ref{tab:scoregroup}, it appears that the number of students who received a grade below 60 remained almost unchanged, even after replacing their original homework grades with the grades generated by ChatGPT. It is worth noting that we only considered the impact of ChatGPT on homework grades in this analysis, as the other exams were conducted in-person and students did not have access to ChatGPT during those exams.



\begin{table}[h]
    \centering
    \begin{tabular}{|c|c|c|c|c|c|}
    \hline
    Mode & 0-20 & 20-40 & 40-60 & 60-80 & 80-100 \\
    \hline
    Original & 8 & 18 & 37 & 131 & 119 \\
    \hline
    Unassisted & 0 &13 & 46 & 134 & 120 \\
    \hline
    Assisted & 0 & 3 & 53 & 134 & 123 \\
    \hline
    \end{tabular}
    \caption{Comparison of the grade distribution. Here the first row represents the grade cut-offs whereas the cells record the number of students within the corresponding grade range.}
    \label{tab:scoregroup}
\end{table}

Besides that, it is important to note that, at present, the reliability of ChatGPT-generated results cannot be fully guaranteed. For instance, our study demonstrates that ChatGPT's overall grade places it at 155 out of 314 students. This suggests that it may not be fully effective in providing assistance to students who outperform it. Conversely, instructors and TAs are capable of offering help to students at different levels, including high-achieving students. Thus, it is important to acknowledge that it has limitations compared to human instructors.

Moreover, it is essential for users to be able to assess the quality of code generated by ChatGPT. While the accuracy of ChatGPT's output is primarily evaluated through test cases in controlled environments, the quality of code must meet more stringent requirements in real-world scenarios. Therefore, programming education remains imperative to equip individuals with the necessary skills to evaluate and improve upon the code produced by language models like ChatGPT. Thus users should at least understand the logic and intention of the code. Therefore, programming education remains vital to developing critical thinking skills and the ability to assess and improve the work produced by language models such as ChatGPT.

On the other hand, students may become overly reliant on the tool instead of developing their own problem-solving skills \cite{hazzan_2023}. Over-reliance on ChatGPT may impact students' critical thinking and problem-solving skills. Moreover, as ChatGPT provides pre-generated responses to problems, students may become less likely to explore alternative solutions or think outside the box. This could lead to a lack of creativity and originality in student work. Moreover, the increasing reliance on ChatGPT for tasks such as code generation raises concerns about the authenticity of the work being produced. In the long run, this may lead to a decline in the quality of education and may diminish the value of computer science degrees.


While ChatGPT can be a valuable tool in computer science education, it also has the potential to negatively impact academic integrity by facilitating plagiarism. With its ability to generate natural language texts and computer programs, ChatGPT can easily be used to produce essays or code that resembles the work of others without proper citation or attribution. Furthermore, if ChatGPT is used for assessment purposes, it may weaken the validity of academic assessments, as it may be difficult to differentiate between the work of a student and that of a machine. This could have serious implications for computer science education, as plagiarism undermines the integrity of academic work and diminishes the value of the educational experience.

Given these concerns, it is crucial to carefully consider how ChatGPT should be integrated into computer science education. Further exploration on how ChatGPT can be used in conjunction with other teaching methods to optimize its advantages while mitigating potential limitations is well-needed \cite{nguyen_2022}. While ChatGPT can undoubtedly be a valuable asset to computer science education, it is imperative to use it responsibly and in tandem with other teaching methods to ensure that students develop a well-rounded set of programming skills. Over-reliance on ChatGPT may negatively impact students' critical thinking and problem-solving abilities, and ethical concerns may arise regarding its potential misuse in areas like plagiarism detection.

Our study offers preliminary evidence that ChatGPT can increase student engagement and enhance learning outcomes in introductory programming courses. Nevertheless, careful consideration must be given to its implementation and use to maximize its benefits while minimizing potential drawbacks.

\section{THREATS TO VALIDITY}
ChatGPT has received a few major updates since its release to improve its performance. Since our empirical results are evaluated on a certain version of ChatGPT, results evaluated on future versions may be different. In addition, because the results produced by ChatGPT largely depend on the prompt, different applications of prompt engineering such as paraphrasing the same question may yield different answers as well. These concerns exist regardless of what kind of tasks we want ChatGPT to perform.
Besides that, it is worth mentioning that results for different programming languages may also vary given the fact they weigh differently in training data. Thus, more popular programming languages such as Python tend to yield better results than less popular programming languages such as Pascal. In short, the differences between our results and the evaluation of ChatGPT performed on other introductory-level programming courses are well-expected.

\section{Conclusion}

In this paper, we share our experience evaluating ChatGPT in an introductory functional programming language course. Our findings show that ChatGPT displays promising results by generating relevant and useful solutions for programming problems. In fact, we discovered that ChatGPT could achieve a perfect score on 16 out of 31 assignment problems without any additional assistance.

Moreover, we conducted experiments to validate the effectiveness of prompt engineering in enhancing ChatGPT's performance on homework and exams. Specifically, we observed that ChatGPT's rank was 220 out of 314 without prompt engineering. However, after implementing prompt engineering, ChatGPT's rank improved significantly to 155. However, ChatGPT presently still struggles with understanding type specifications and inferring the type of an expression. It also struggles with larger programming tasks such as implementing an interpreter. 
Building on the various merits of ChatGPT, we delve into how this technology can enhance students' learning experience. One of the key benefits is the ability to provide real-time support, which can greatly assist learners in overcoming any obstacles they encounter during the course. Furthermore, we also highlight potential directions that instructors can explore to leverage ChatGPT's capabilities. For instance, ChatGPT can generate lesson plans tailored to the specific needs of individual learners. This feature can enable instructors to personalize the learning process and optimize their students' progress.


We believe that ChatGPT has the potential to revolutionize computer science education, but there are many concerns that must be addressed before widespread adoption. These include the possibility of generating code of insufficient quality, over-reliance on technology, and potential issues with academic integrity. Therefore, further exploration is necessary to investigate how language models such as ChatGPT can be integrated into the current education system. This involves determining how the technology can be utilized to its full potential while minimizing the risks of its application. In summary, while we recognize the potential of ChatGPT to transform computer science education, careful consideration must be given to the challenges associated with its use. 

\begin{acks}
This work was supported, in part, by Individual Discovery Grants from the Natural Sciences and Engineering Research Council of Canada, the Canada CIFAR AI Chair Program, and Social Sciences and Humanities Research Council (SSHRC). 
\end{acks}

\newpage
\bibliographystyle{ACM-Reference-Format}
\bibliography{sample-base}

\appendix

\end{document}